\documentclass[aps,prc,showpacs,twocolumn,superscriptaddress, amssymb]{revtex4}
\usepackage{color}
\usepackage{graphicx}

\begin{document}

\title{Inclusive pion double charge exchange on $^{16}$O 
above the delta resonance}
\author{A.P. Krutenkova}
\email[]{Anna.Krutenkova@itep.ru}
\affiliation{Institute of Theoretical and Experimental Physics, Moscow 117218,
Russia}
\author{T. Watanabe}
\affiliation{Department of Physics, Tohoku University, Sendai 980-8578, Japan}
\author{D. Abe}
\affiliation{Department of Physics, Tohoku University, Sendai 980-8578, Japan}
\author{Y. Fujii}
\affiliation{Department of Physics, Tohoku University, Sendai 980-8578, Japan}
\author{O. Hashimoto}
\affiliation{Department of Physics, Tohoku University, Sendai 980-8578, Japan}
\author{V.V. Kulikov}
\affiliation{Institute of Theoretical and Experimental Physics, Moscow 117218,
Russia}
\author{T. Nagae}
\affiliation{High Energy Accelerator Research Organization (KEK), Tsukuba,
Ibaraki 305-0801, Japan}
\author{M. Nakamura}
\affiliation{Graduate School of Science, University of Tokyo, 
Tokyo 113-0033, Japan}
\author{H. Noumi}
\affiliation
{High Energy Accelerator Research Organization (KEK), 
Tsukuba, Ibaraki 305-0801, Japan}
\author{H. Outa}
\affiliation
{High Energy Accelerator Research Organization (KEK), 
Tsukuba, Ibaraki 305-0801, Japan}
\author{P.K. Saha}
\affiliation{Laboratory of Physics
Electro-Communication University, 
Neyagawa, Osaka 572-8530,
Japan}
\affiliation{Japan Atomic Energy Research Institute, Tokai, Ibaraki
319-1195, Japan}
\author{T. Takahashi}
\affiliation{Department of Physics, Tohoku University, Sendai 980-8578, Japan}
\author{H. Tamura}
\affiliation{Department of Physics, Tohoku University, Sendai 980-8578, Japan}

\date{\today}

\begin{abstract}

The forward inclusive pion double charge exchange reaction,
$^{\mathbf{16}}$O$(\pi^-,\pi^+)$X, at $T_0=$ 0.50 and 0.75 GeV
has been studied in the kinematic region where an additional pion
production is forbidden by energy-momentum conservation.
The experiment was performed with the SKS spectrometer at KEK PS. 
The measured ratio of double charge exchange cross-section for these energies
$\langle$d$\sigma$(0.50 GeV)/d$\Omega \rangle$ /
$\langle$d$\sigma$(0.75 GeV)/d$\Omega \rangle$
 = 1.7 $\pm$ 0.2, disagrees with the value of 7.2 predicted within
the conventional sequential single charge exchange mechanism.
Possible reasons for the disagreement are  discussed in connection 
with the Glauber inelastic rescatterings.

\end{abstract}

\pacs{13.75.Gx, 25.10.+s, 25.80.Gn}

\keywords{}

\maketitle

\begin{figure}[bhtp]
\includegraphics[bb=0 0 567 580,width=8.5cm]{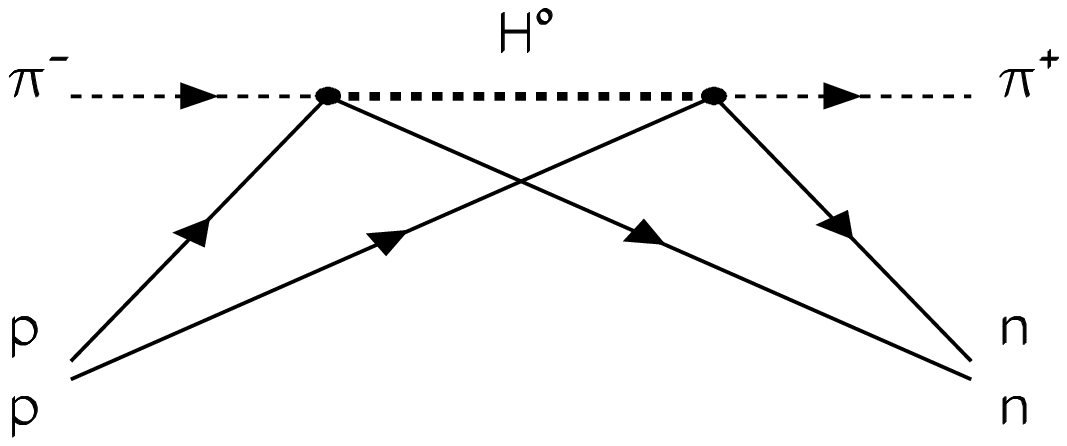}
\vspace*{-6.5cm}
\caption[] {Diagrams contributing to pion double charge exchange
on a nucleus:\\ (a) sequential single charge exchange (SSCX) with $\pi^0$ 
in the intermediate state (H$^0$ = $\pi^0$) and \\(b) 
inelastic Glauber rescatterings with two pions in the
intermediate state (H$^0$ = $\pi^+$$\pi^-$ and $\pi^0$$\pi^0$) .}
\label{diagram1}
\end{figure}

The pion double charge exchange reaction (DCX) is a good testing ground for 
probing nucleon-nucleon correlations in a nucleus due to its two nucleon 
nature. 
In the past this reaction was extensively studied at LAMPF energies 
$T_0 \leq$ 0.50 GeV \cite{johnson} and it was shown that 
the DCX can be described reasonably well in the framework of sequential 
single charge exchange (SSCX)  models. In these models, pion DCX
is explained by two successive single charge exchanges with a real neutral 
pion in an intermediate state (Fig.~\ref{diagram1}(a), H$^0$ = $\pi^0$).
The SSCX mechanism predicts rapid decrease of 
the forward DCX cross sections 
for pion energies above 0.5 GeV.
For this reason high-energy DCX was suggested (see, e.g. 
Ref.~\cite{hashimoto,oset}) 
as a probe of the short-range nucleon-nucleon correlations in a nucleus and 
new mechanisms of pion propagation in the nuclear medium.

Pion DCX above 0.5 GeV can be studied experimentally only in the narrow 
kinematic region of outgoing pion momenta near the high energy end point 
where production of additional pions is forbidden by the energy-momentum 
conservation. This requires both a high intensity pion beam and a good 
spectrometer system in the 1 GeV/c region. We studied the process 
  \begin{equation}
    \pi^- +{}^{16}\mathrm{O} \rightarrow \pi^{+} + \mathrm{X}
  \end{equation}
  using superconducting kaon spectrometer (SKS) \cite{sks} 
  at KEK 12-GeV PS sharing the apparatus of the E438 experiment which measured 
  the ($\pi^-$,K$^+$) reaction \cite{noumi}.
  In this paper we give a brief description of the experimental setup, 
  data taking and analysis procedures which are discussed in detail 
  in  Ref.~\cite{pranab}.

Negative pions of kinetic energy $T_0=$ 0.50 and 0.75 GeV with the beam flux
of (1 -- 2) $\times$10$^6$ pions per spill were delivered 
to the target at the K6 beamline.
The beam spectrometer consisted of a $QQDQQ$ magnet system, four sets of
the drift chambers with 24 layers of sense wire planes 
in total used for the momentum reconstruction and three trigger counters: 
a freon-gas $\mathrm{\breve{C}}$erenkov counter ($GC$) and two sets of 
segmented scintillation counters $BH1$ and $BH2$. 
Beam particle, 
identified by the beam spectrometer, interacted in the  5-cm 
long H$_2$O target. Positive particles emitted in the forward direction 
(the reaction angle was $\theta \leq 15^o$) in the process
\begin{equation}
\pi^- +  A \rightarrow \mathrm{(e^+}, \pi^+, \mathrm{p}) +  \mathrm{X}
\end{equation}
  were measured with the SKS spectrometer consisting of a
  superconducting dipole magnet, four sets of the drift chambers 
  with 22 layers of sense wire planes in total and several 
  trigger counters:
  $TOF$ wall and Lucite $\mathrm{\breve{C}}$erenkov counter wall ($LC$) 
  for proton suppression. Data were taken with a trigger 
  $BH1 \times BH2 \times  \overline{GC} \times TOF \times LC$. 
  The trigger rates varied from 60 to 400 per spill depending on the 
  pion beam energy. At each beam energy two settings of the magnetic field 
  ($I_{SKS}$ = 145 A and 175 A at $T_0=$ 0.50 GeV and 
  $I_{SKS}$ = 272 A and 320 A at $T_0=$ 0.75 GeV) 
  were used to cover the wider range of the outgoing pion energy $T$.
  These pions have to be discriminated from protons and positrons.
  Time-of-flight measurement between $BH2$ and the $TOF$ wall was used 
  to reject protons.
 
The beam flux used and the numbers of selected events  
are given in the Table I.
The events have passed the standard SKS cuts on 
the number of hits in the drift chambers,
on the vertex position in the target, etc.
The selection procedure is described in detail in Ref.~\cite{pranab}.
We used additional cut on the reaction angle $4^o \leq \theta \leq 6^o$
which was applied to reduce the positron background as described below
and to facilitate a comparison with the existing data ~\cite{bur,npa}.
Apart from the total number of selected events, $N_{tot}$, 
the numbers of events in energy 
ranges 0 $\leq \Delta T = T_0 - T \leq$ 80(140) MeV, $N_{80}$ ($N_{140}$)
are given.

\begin{table}[bhpt]
\caption{The numbers of selected events used for cross section calculations
for different energy and SKS current settings.  $N_{\pi^-}$ is a beam flux
in units of $10^{9}$.}
\begin{center}
\begin{tabular}{|l|l|l|l|l|l|}
$T_0$, GeV& $I_{SKS}$, A& $N_{\pi^-}$ &  $N_{tot}$ &$N_{80}$ &  $N_{140}$ \\ 
\hline
 0.5  & 145 &  7.2 & 1599 &    197 & 1033    \\[0.10cm]
 0.5  & 175 & 15.7 & 1017 &    433 &   --    \\[0.10cm]
 0.75 & 272 & 25.2 & 7710 &    362 & 1661    \\[0.10cm]
 0.75 & 320 & 32.6 & 4859 &    621 & 2449    \\[0.10cm]
\end{tabular}
\end{center}
\end{table}
 
Among the detected particles there is a sizable fraction of positrons 
which originate from two sources. Electrons 
  contaminating the beam can radiate the photons, photon conversion in the 
  target produces the positron. Positrons can also result from the single 
  charge exchange $\pi^- \rightarrow \pi^0$ processes in the 
  target followed by the $\pi^0 \rightarrow \gamma \gamma$ decays and photon 
  conversion or direct Dalitz decays of the produced $\pi^0$'s.

\begin{figure}[bhtp]
\includegraphics[bb=0 0 567 660,width=7cm]{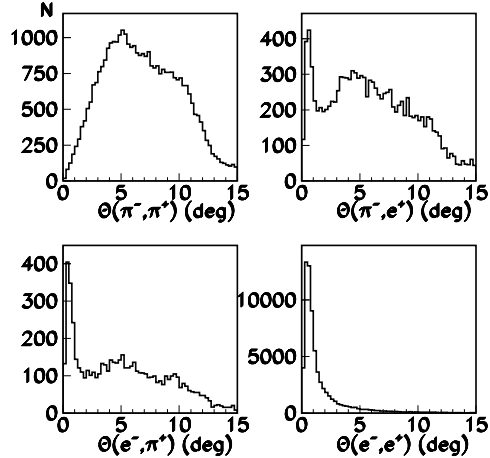}
\vspace*{-2.2cm} 
\caption [] {The angular distributions for the
reactions ($\pi^-,\pi^+$), ($e^-,\pi^+$), ($\pi^-,e^+$) and
($e^-,e^+$) at  $T_0 =$ 0.75 GeV/c and $I_{SKS} = $ 320 A as an
example. Detection efficiencies of $GC$ and $AEC$ measured in the beam were
$\epsilon ^{GC}$ = 0.94 (0.095) and $\epsilon ^{AEC}$ = 0.87 (0.13)
for electrons (pions). 
} 
\label{diagram13}
\end{figure}

Special run was devoted to study the positron background. 
An additional  aerogel ($n$ = 1.01)
  $\mathrm{\breve{C}}$erenkov counter ($AEC$) was placed downstream of the 
  target. 
It was tested in the 150 MeV electron 
  beam from the electron LINAC at Tohoku University. 
  The signals from the $GC$ and the $AEC$ $\mathrm{\breve{C}}$erenkov counters 
  were measured and used off-line to separate pions from electrons/positrons.
  Such off-line analysis allowed to disentangle the 
  following four reactions: ($\pi^-,\pi^+$), ($e^-,\pi^+$), ($\pi^-,e^+$) and 
  ($e^-,e^+$)
  registered with the $BH1 \times BH2 \times TOF \times LC$ trigger.  
  Angular distributions for these four reactions are presented in  
  Fig.~\ref{diagram13} for $T_0 = $ 0.75 GeV/c, as an example. 
  The reaction ($e^-,e^+$) gives a sharp peak of the 
  electromagnetic nature at 0$^o$. 
  For this reason, only the data in the angular 
  region of $4^o \leq \theta \leq 6^o$ were used 
  for further analysis.    
By solving the set of four linear equations the correction factor  
for the DCX cross section
$R = N(\pi^-,e^+)_{true}/ (N(\pi^-,e^+)_{true} + N(\pi^-,\pi^+)_{true}$)
was calculated using the $GC$ and the $AEC$ efficiencies measured in the beam. 
For the kinematic region of pion DCX 
($\Delta T = T_0 - T \leq m_{\pi} \approx 140$ MeV) it
was found to be 0.54 $\pm$ 0.08  
(0.35 $\pm$ 0.06) for $T_0$ = 0.50 (0.75) GeV. 
Error in $R$ is dominated by statistics in the chosen kinematic region. 

We determined the differential cross section of the reaction (1) 
as a function of $\Delta T$, 
corrected for ionization energy loss of initial and final pions 
  in the target. The angular acceptance 
  was calculated using Monte Carlo simulation which took into 
  account the detector geometry, map of the magnetic field and the 
  selection criteria applied in the analysis. The cross section was 
  corrected for 
  the muon contamination in the beam, detector efficiencies, efficiencies of 
  the analysis and pion decays just in a similar way as it has been done
  in \cite{pranab}. 
  The product of the correction factors was calculated to be $0.320 \pm 0.005$
  for $T_0$ = 0.50 GeV and $0.304 \pm 0.004$ for $T_0$ = 0.75 GeV. 
  The positron background was taken into account using the $R$ value which 
  within the experimental errors did not depend on $\Delta T$. 
  To calibrate the $\Delta T$ scale and energy resolution we used 
  $\Sigma^-$ hyperon peak reconstructed in the reaction 
  $\pi^- $p$ \rightarrow $K$^+ {\Sigma^-}$ on the CH$_2$ target 
  and found the energy resolution to be less than 3 MeV (FWHM) 
  (see \cite{noumi,pranab}).  The systematic errors (without the uncertainty
  in the $R$ value) which include 
  the uncertainty of $\Delta T$-scale 
  calibration are estimated to be less than 10\%.

\begin{figure}[bhtp]
\includegraphics[bb=0 0 567 660,width=8.5cm]{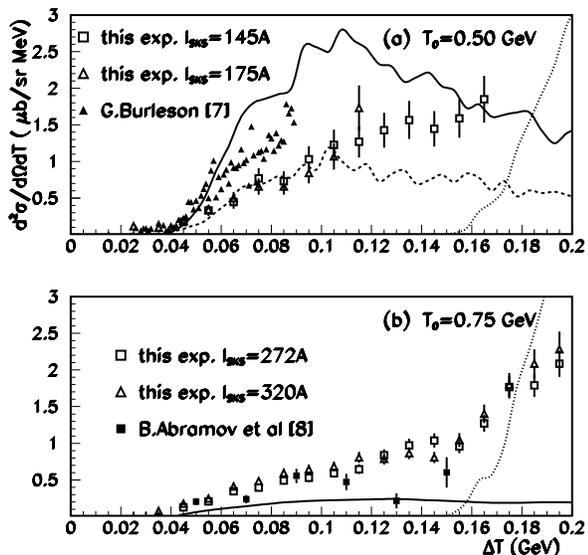}
\vspace*{-2.8cm} 
\caption [] {Differential cross section for
$\pi^-  + {}^{16}$O$ \rightarrow \pi^+ + $X at $\theta = 4 - 6^o$
as a function of $\Delta T = T_0 - T$ for different pion beam
kinetic energies : (a) $T_0=0.50$ GeV; (b) $T_0$ = 0.75 GeV. Curves are 
described in the text.
}
\label{diagram17} 
\end{figure}

  The differential cross section of the reaction (1) for 
  $4^o \leq \theta \leq 6^o$
  at $T_0=$ 0.50 and 0.75 GeV calculated for two settings of the magnetic 
  field at each energy is shown in Fig.~\ref{diagram17} where only statistical 
  errors are given. The results for different SKS currents 
  agree within the statistical errors in the overlap regions. 

  In Fig.~\ref{diagram17} we compare our results 
  to  the available data from LAMPF for the reaction 
  $\pi^+ +{}^{16}\mathrm{O} \rightarrow \pi^{-} + \mathrm{X}$ (3) 
  at $T_0=$ 0.50 GeV (a) \cite{bur} and 
  from ITEP at $T_0=$ 0.75 GeV (b) \cite{npa}. 
  The results of the theoretical calculations are 
  also shown in the figures:
  solid and dashed curves correspond to the calculations within SSCX model 
  (see Ref.~\cite{npa},~\cite{vacas}) without and with a core polarization  
  ~\cite{oset} and the dotted curve corresponds to the cascade type calculation 
  from \cite{vacas}.
  The measured cross sections grow with $\Delta T$. 
  Fig.~\ref{diagram17}(b) shows that the slope of the $\Delta T$ 
  dependence changes around 150 MeV, which is near the threshold  
  ($\Delta T \simeq$  140 MeV) of additional pion production 
  $\pi^- + ^{16}$O  $\rightarrow \pi^+ + \pi^- +$ X~.
  For $\Delta T$ below 140 MeV this process is forbidden by the 
  energy-momentum conservation. The cascade type calculation does not 
  explain correctly the rise of the cross section above $\Delta T \simeq$ 
  140 MeV. This is not surprising as there are no direct measurements of the 
  pion production cross section in this energy region and the calculation
  is based on the far extrapolation of the existing data. 
  At $T_0$ = 0.50 GeV our measurements are systematically lower than the
  data from LAMPF, the reason for that is not yet understood. 
  At $\Delta T \leq$ 0.12 GeV our results agree with the SSCX calculations 
  with the core polarization (dashed curve). 
  At $T_0=$ 0.75 GeV, however, the picture is somewhat different. Our 
  results are in reasonable agreement with the previous measurements 
  from ITEP \cite{npa} which had larger errors and both sets of data are 
  systematically above the SSCX predictions.
  The cross section $\langle$d$\sigma$/d$\Omega \rangle_{80}$, 
  integrated over the $\Delta T$ from 0 to 80 MeV,
  and $\langle$d$\sigma$/d$\Omega \rangle_{140}$, integrated from 0 to 140 MeV,
  are presented in Table II. 
  The errors quoted in the table include the statistical errors and 
  the uncertainties coming from the determination of $R$.
  For the 80 MeV range
  they are comparable while for the 140 MeV one 
  the second source dominates.
  The measurements at different SKS currents are in good agreement
  and were averaged.

\begin{figure}[bhtp]
\includegraphics[bb=0 0 567 640,width=8.5cm]{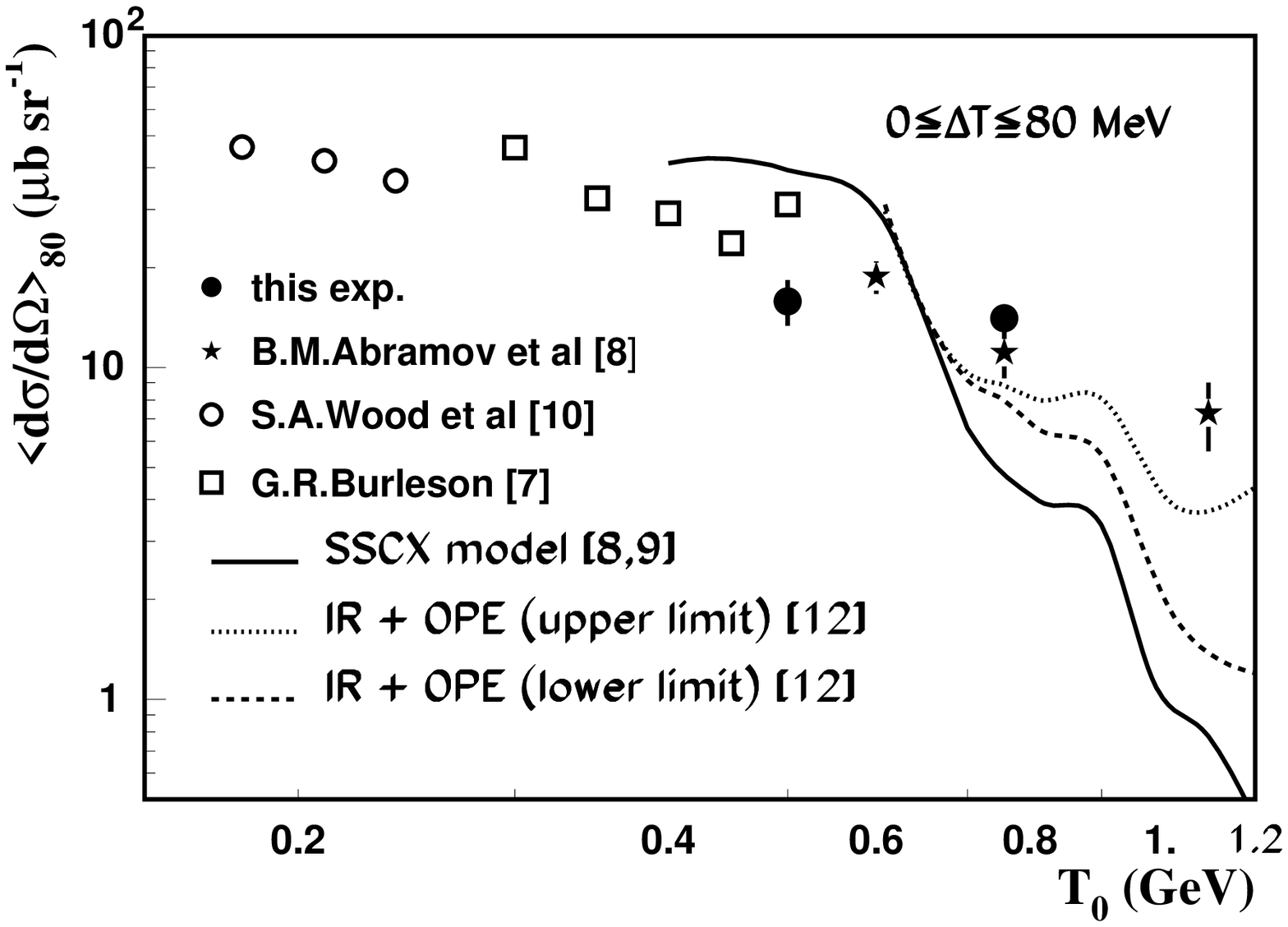}
\vspace*{-3.0cm}
\caption [] { Energy dependence of the DCX cross
section integrated over the $\Delta T$ range from 0 to 80 MeV. }
\label{diagram18}
\end{figure}

\begin{figure}[bhtp]
\includegraphics[bb=0 0 567 600,width=8.5cm]{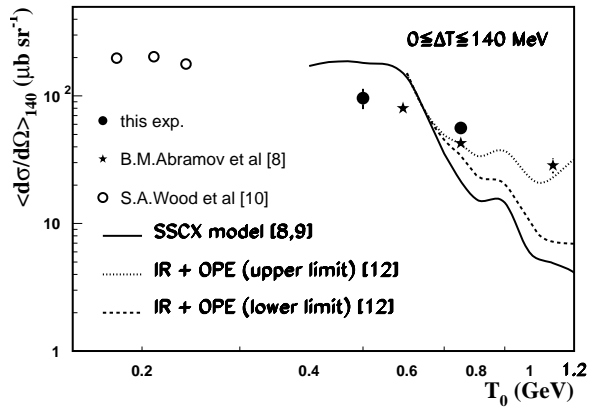}
\vspace*{-3.0cm}
\caption [] {Energy dependence of the DCX cross
section integrated over the $\Delta T$ range from 0 to 140 MeV. }
\label{diagram19}
\end{figure}

\begin{table}[bhpt]
\caption{DCX cross sections integrated over the  
$\Delta T$ ranges from 0 to 80 MeV
($\langle$d$\sigma$/d$\Omega \rangle_{80}$) and from 0 to 140 MeV 
($\langle$d$\sigma$/d$\Omega \rangle_{140}$) with their statistical errors 
and errors originating from the values of $R$.}
\begin{center}
\begin{tabular}{l|cc|cc}
 & \multicolumn{2}{c|}{$\langle$d$\sigma$/d$\Omega \rangle_{80}$, $\mu
b/sr$} & \multicolumn{2}{c}{
$\langle$d$\sigma$/d$\Omega \rangle_{140}$, $\mu b/sr$}  \\[0.15cm]
\hline
$T_0,$ GeV  &0.50 &0.75  &0.50  &0.75 
\\[0.15cm]\hline
$I_{SKS}$ = 145/272 A
& 15.1$\pm$3.5   & 12.7$\pm$1.6 & 96.2$\pm$17.5  & 53.2$\pm$5.2   \\[0.25cm]
$I_{SKS}$ = 175/320 A
& 16.6$\pm$3.6 &15.5$\pm$1.8 &  &59.0$\pm$5.7
\\[0.15cm]\hline
$$
Averaged &15.9$\pm$3.2&14.1$\pm$1.5
&96.2$\pm$17.5    &56.1$\pm$5.4\\[0.25cm]
\end{tabular}
\end{center}
\end{table}
  The energy dependence of the integrated cross sections 
  $\langle$d$\sigma$/d$\Omega \rangle_{80}$ and 
  $\langle$d$\sigma$/d$\Omega \rangle_{140}$ is shown in
  Fig.~\ref{diagram18} and Fig.~\ref{diagram19}. 
  Also shown are the data of \cite{wood} at $T_0 = $ 0.18, 
  0.21 and 0.24 GeV for $\theta = 25^o$, the results of \cite{bur} 
  for the reaction (3) and the results of \cite{npa} 
  at 0.6 -- 1.1 GeV.


The ratios $\langle$d$\sigma$(0.5 GeV)/d$\Omega \rangle_{80}$/
$\langle$d$\sigma$(0.75 GeV)/d$\Omega \rangle_{80}$
 = 1.1 $\pm$ 0.2 and
$\langle$d$\sigma$(0.5 GeV)/d$\Omega \rangle_{140}$/
$\langle$d$\sigma$(0.75 GeV)/d$\Omega \rangle_{140}$
 = 1.7 $\pm$ 0.2
are much smaller than SSCX predictions which are 7.5 and 7.2 
for 0-80 MeV and 0-140 MeV $\Delta T$ ranges respectively.
  Our results, based on large statistics, strongly support 
  an  observation of ITEP \cite{npa} of the anomalous 
  energy dependence of inclusive pion DCX cross section in the beam
  energy range of 0.6 - 1.1 GeV . The observed dependence suggests 
  that either in this energy range we have a contribution of a new DCX 
  mechanism or the charge exchange $\pi$N amplitude is modified in a 
  nuclear medium.

It was shown in Ref.~\cite{kaid} that the virtual multipion intermediate 
states, the so-called Glauber inelastic rescatterings (IR), 
seem to be important for the description of the inclusive DCX cross 
section at energies $T_0 \gtrsim $ 0.6 GeV. In \cite{kakru} the cross 
section of the reaction (1) was taken as the sum of two contributions,
\footnote{Contributions from $\eta ^0$ in diagram of Fig.~\ref{diagram1} 
    (H$^0$ = $\eta^0$) and from the mechanism of meson exchange currents 
    appear to be small (see, respectively,~\cite{kaid} and \cite{luis}).}
one with an intermediate $\pi^0$ (SSCX, Fig.~\ref{diagram1}(a), 
H$^0$ = $\pi^0$) and another one with an intermediate $2\pi$ state 
(IR mechanism, Fig.~\ref{diagram1}(b), H$^0$ = $\pi^+$$\pi^-$ 
and $\pi^0$$\pi^0$). The IR contribution was estimated in the framework 
of the Gribov-Glauber approach to DCX within the OPE model. 
  The dotted and dashed curves in Figs.~\ref{diagram18} and \ref{diagram19} 
  correspond to the upper and the lower limits of the theoretical estimates of 
  \cite{kaid}. 
  The dotted curve is much closer to the experimental data, especially 
  for $\langle$d$\sigma$/d$\Omega \rangle_{140}$, which represents a 
  considerable improvement over the SSCX calculation.

It is worth mentioning here that IR is a new mechanism of pion 
propagation in a nuclear medium totally absent in cascade models 
where real pions and on-shell amplitudes are always used.

  Modification of the charge exchange $\pi$N amplitude in a nuclear 
  medium can be calculated using the approach developed in \cite{kondrat} 
  for $\gamma$N interactions. Within this approach modification of a $\pi N$ 
  or $\gamma N$ amplitude in a nuclear medium is explained by the widening 
  of the baryon isobars due to the new channels like NN$^* \rightarrow$ NN 
  being open in a nucleus. However the detailed calculations within this 
  approach so far have not been performed.

  In conclusion, we have measured the cross section of the inclusive DCX 
  reaction  $\pi^- + {}^{16}\mathrm{O} \rightarrow \pi^{+} +$ {X} 
  at two energies $T_0 =$ 0.50 and 0.75 GeV and found the ratio of the 
  measured cross sections $\langle$d$\sigma$(0.5 GeV)/d$\Omega \rangle_{140}$/
  $\langle$d$\sigma$(0.75 GeV)/d$\Omega \rangle_{140}$ = 1.7 $\pm$ 0.2
  to be much lower than the value of 7.2 predicted by the conventional SSCX 
  mechanism. Evidently new mechanisms are needed to explain this discrepancy and
  the mechanism of inelastic rescatterings considered in \cite{kaid} seems to 
  be one of the good candidates.
\begin{acknowledgments}
We are grateful to the staff of the SKS spectrometer for the assistance 
during the runs. We thank the ITEP Scientific Director M.V. Danilov for 
the support of Russian participation in the experiment, A.Williams for 
providing the experimental data used in  Fig.~\ref{diagram17}(a) and E.Oset 
for stimulating discussions. We are also grateful to Dr. P.A.Murat and Dr. R.J.Schneider for
reading the manuscript. A.P.K. thanks Tohoku University for the 
hospitality. This work was supported in part by Grant FTsNTP-40.052.1.1.1113.
\end{acknowledgments}


\bibliography{basename of .bib file}

\end{document}